\documentclass[prl,twocolumn,notitlepage,amsfonts,amssymb,amsmath,floatfix,tightenlines,superscriptaddress]{revtex4-2}

\usepackage{amsfonts,amssymb}
\usepackage{amsmath,times}
\usepackage{graphicx}
\usepackage[colorlinks=true,linkcolor=blue,anchorcolor=red,citecolor=blue,urlcolor=blue]{hyperref}
\usepackage{color}
\usepackage{cleveref}
\usepackage[lofdepth,lotdepth,caption=false]{subfig}
\usepackage{soul,xcolor}
\usepackage{float}      

\begin{document}

\title{Quench induced collective excitations: from breathing to acoustic modes}

\author{Shicong Song}
\affiliation{Department of Physics, Florida Atlantic University, 777 Glades Road, Boca Raton, FL 33431-0991, USA}

\author{Ke Wang}
\affiliation{Department of Physics and James Franck Institute, University of Chicago, Chicago, Illinois 60637, USA}

\author{Zhengli Wu}
\affiliation{Department of Physics, University of Oxford, Oxford, OX1 3PU, UK}

\author{Andreas Glatz}
\affiliation{Materials Science Division, Argonne National Laboratory, 9700 S. Cass Avenue, Argonne, Illinois 60439, USA}
\affiliation{Department of Physics, Northern Illinois University, DeKalb, Illinois 60115, USA}

\author{K. Levin}
\affiliation{Department of Physics and James Franck Institute, University of Chicago, Chicago, Illinois 60637, USA}

\author{Han Fu}
\affiliation{Department of Physics, Florida Atlantic University, 777 Glades Road, Boca Raton, FL 33431-0991, USA}

\begin{abstract}
In trapped Bose-Einstein condensates, interaction quenches which are abrupt changes of the interaction strength typically implemented via Feshbach tuning, are a practical and widely used protocol to address far-from-equilibrium collective modes.
Using both numerical Gross Pitaevskii and analytical schemes we study these interaction-quench-induced collective modes
in a harmonically trapped two-dimensional Bose--Einstein condensate 
contrasting the behavior found at low and high energies.
In the low-lying regime, we characterize realistic circumstances in which there is a breakdown of the expected
scale invariance so that the collective excitations follow hydrodynamic theory instead of the predictions given by SO(2,1)
conformal symmetry.
In the high energy regime, we focus on important trap effects associated with acoustic oscillations which
have been of interest experimentally.
This comprehensive analysis of the collective excitations in trapped two-dimensional Bose-Einstein condensates
is experimentally accessible. Through their frequencies and damping, this reflects an important built-in spectroscopy of such many-body states.
\end{abstract}

\date{\today}

\maketitle

\section{Introduction}
Understanding the dynamics of quantum many-body systems far from equilibrium remains a central challenge in condensed-matter and atomic physics \cite{RevModPhys.83.863, LAMACRAFT2012177}. Ultracold atomic gases, particularly Bose--Einstein condensates (BECs), have emerged as highly tunable platforms for probing such dynamics, with relevance to quantum simulation, hydrodynamics, and collective-mode engineering~\cite{bloch2012quantum, PhysRevA.109.013316,NaturePhys8267}. Of particular interest are
the collective modes of these atomic gases which provide crucial information about the trapped condensate state\cite{Griffin_1993, RevModPhys.80.1215, 2016becs.book.....P}. 

In this paper we address these collective modes which are stimulated by a quench in a harmonically trapped
two-dimensional (2D) Bose-Einstein condensate (BEC). We
note that theory tends to incorporate an idealized description
often assuming a homogeneous BEC, yet in practice the gas
is confined by traps that introduce intrinsic inhomogeneity. Additionally the interaction potential in an idealized
situation is modeled by a contact interaction. Experimentally,
one approaches this contact limit by preparing a sufficiently
dilute atomic gas. This may not apply in general, as for example, in the presence of narrow Feshbach resonances \cite{Nature.392.151(1998), RevModPhys.82.1225}. Additionally one has to control the two dimensionality and confine consideration to temperatures well away from the BKT
transition as at this temperature there will be changes in the
compressibility, the equation of state and viscosities leading
to departures from an idealized description of the collective
modes \cite{JMKosterlitz_1973, PhysRevA.66.043608}. 

Our goal here is to address the collective modes under more
realistic circumstances which are outlined above. We focus
our attention separately on the two extreme regimes of 2D
gases by studying the collective modes associated with both
high and low momentum $k$.The current work employs analytical and numerical methodologies to comprehensively understand the non-equilibrium dynamics
of trapped BECs. The BEC dynamics are modeled using the Gross-Pitaevskii (GP) equation within a harmonic trap: 
\begin{equation} \label{GPE_latex}
i\hbar\frac{\partial\psi}{\partial t} = \left[ -\frac{\hbar^{2}}{2m}\nabla^{2} +\frac{1}{2}m\omega_0^2r^2 +g_{0}\left|\psi\right|^{2} -\mu \right]\psi + \chi(\mathbf{r},t),
\end{equation}
Here $g_0$ represents the interaction strength, $\omega_0$ is the trap frequency, and $\mu$ is the chemical potential. We numerically add noise seeds $\chi(\mathbf{r},t)$ to simulate quantum fluctuations \footnote{A detailed description of our numerical methodology is provided in Appendix~\ref{SeedingQ}.}. We study the BEC in the Thomas-Fermi regime where $\mu/\hbar\omega_0\gg1$ with Thomas-Fermi radius $r_0=\sqrt{2\mu/m\omega_0^2}$~\cite{PethickSmith2008,PitaevskiiStringari2003}. 

The low-lying excitations in such a system have been addressed using
a hydrodynamic approach developed by Stringari \cite{PhysRevLett.77.2360, RevModPhys.71.463}, and 
a SO(2,1) conformal symmetry-based scheme which derives from scale invariance \cite{PhysRevA.55.R853}). While based on different assumptions, these
theories both arrive at the same frequency for the lowest mode in the sequence.
Experiments have confirmed this frequency~\cite{Bloch}
where they unfortunately, cannot be readily differentiated.
However, the higher excitation spectrum yields distinct behaviors. Understanding
and verifying these differences is of particular interest here, theoretically, 
and in future experiments.

Our more detailed studies show a break-down of scale invariance associated with short observational length scales (resulting from the finite cut-off inherent in experiments and numerical simulation). What one sees often is that two types of characteristic frequencies appear. Breathing modes emerge that deviate from the Pitaevskii's even-integer sequence $\omega_n = 2n\omega_0$ \cite{PhysRevA.55.R853}, yet are captured by Stringari's hydrodynamic analysis \cite{PhysRevLett.77.2360}.
Our study shows that by tuning the quench strength, this deviation can be more pronounced.

In the regime of higher energy and momentum $k$ 
the collective modes cross
over to a more conventional
sound mode description deriving from Bogoliubov theory.
Here momentum becomes (approximately) a good quantum number, thus replacing the numerical excitation mode index. 
While this regime has
received attention experimentally \cite{ChengChin}, what has not been addressed in detailed theory is the important role of the confining
trap.

Our systematic, numerical Gross Pitaevskii calculations show that the leading effect of harmonic confinement can be absorbed into a renormalized global chemical potential, yielding a trap-modified Bogoliubov dispersion that reconciles simulations with measurements \cite{ChengChin}. It
addresses discrepancies in Ref.~\onlinecite{ChengChin} between previous theory and experiment. Our findings also clarify the observed decay and 
finite lifetime of these oscillations in Ref.~\onlinecite{ChengChin}. 

\section{Investigating low-lying excitations}
Over the past decades, the study of Bose-Einstein condensates confined in harmonic traps has received attention, particularly focused on the excitations of low-$k$. At the quantum level, a fundamental structure was uncovered by Pitaevskii and Rosch, who demonstrated the existence of a hidden conformal symmetry in a 2D BEC that persists even in the presence of harmonic confinement \cite{PhysRevA.55.R853}. 
More recently, Maki and Zhou \cite{PhysRevA.109.L051303} discovered an emergent conformal dynamics even in one-dimensional systems where an intrinsic scale symmetry is absent.

This interesting symmetry, arising from the algebraic structure of SO(2,1) and deriving from the Hamiltonian for 2D BECs, guarantees the existence of universal oscillations with a frequency precisely equal to $2n\omega_0$. 
The significance of this algebraic structure is that the Hilbert space naturally decomposes into irreducible representations, leading to an inherently simpler energy spectrum for all eigenstates of the form $E_n = E_0 + 2n\hbar\omega_0$, where $n$ is given by positive integers.

However, this theoretical result depends sensitively on the form of the interaction potential. In the 2D case, an on-site interaction can be modeled by a contact interaction potential ``$g\delta^2(r)$'', which successfully preserves scale invariance and results in conformal symmetry of the system. Nevertheless, the assumption of a precisely on-site interaction is only valid at length scales much larger than the scattering length. It also fails in the neighborhood of a narrow Feshbach resonance.

Our numerical simulations, (like in experiments which inevitably involve a cut-off length for the idealized on-site interaction),
also naturally impose a cut-off due to discretizations when implementing the continuous model. This, thereby, allows us to study the system in a way that simulates the laboratory environment.

\begin{figure}[h]
\centering
\includegraphics[width=\linewidth]{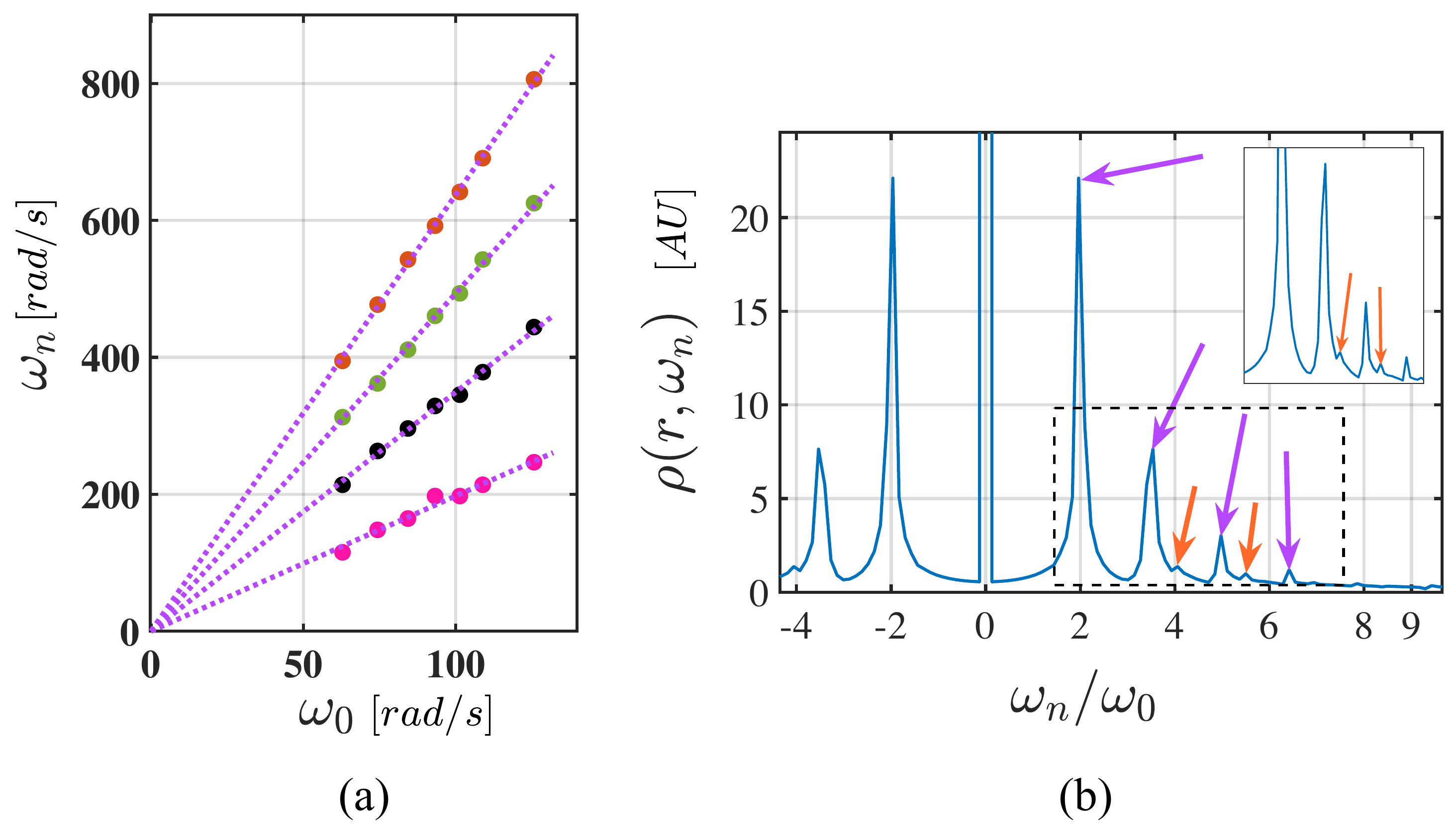}
\caption{
\textbf{(a)} Scaling of the collective-mode frequencies of a trapped condensate with trap frequency $\omega_0$. The panel plots the extracted mode frequencies $\omega_n$ ($n=1,2,3,4$ and $l=0$), obtained from the spectral peaks of $\rho(r,\omega_n)$ at $r=10$ grid points, as $\omega_0$ is varied from $60 ~\mathrm{rad/s}$ to 
$125~\mathrm{rad/s}$. A linear regression yields $\omega_n/\omega_0 = 1.98,\,3.49,\,4.94$ and $6.38$, in close agreement with the theoretical predictions $2.00,\,3.46,\,4.90,$ and $6.32$ from Eq.~\ref{Sfrequency}.\\    
\textbf{(b)} Spectral distribution for $\omega_0 = 125~\mathrm{rad/s}$, where the lowest four dominant peaks (purple arrows) mark the collective modes, and additional minor peaks (orange arrows) are associated with conformal-symmetry modes which are suppressed and deformed at this $r$.}
\label{Stringari_mode}
\end{figure}

\begin{figure}[h]
\raggedright
\hspace*{-4mm}
\includegraphics[width=1.0\linewidth]{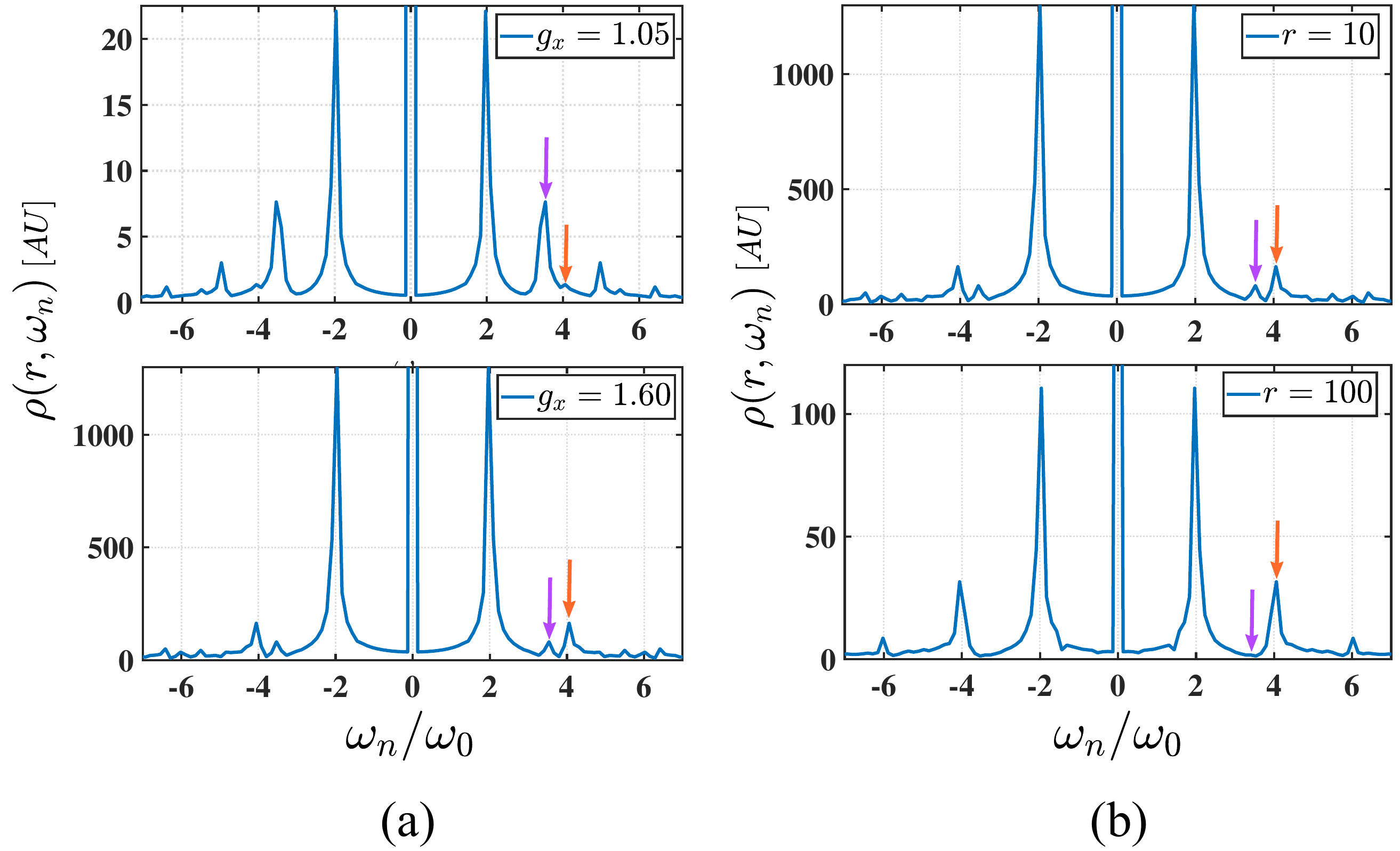}
\caption{
\textbf{(a)} Dependence of the spectral response $\rho(r,\omega_n)$ on quench strength at a fixed short observation length scale $r$. The two panels show results for $g_x = 1.0$ and $g_x = 1.6$. As the quench strength increases, the conformal mode contribution (orange arrows) becomes increasingly dominant relative to the hydrodynamical 
modes (purple arrows).\\
\textbf{(b)} Dependence on observational length scale for the same strong quench ($g_x = 1.6$). The two panels correspond to different $r$ values. At larger lengths $r$, the conformal modes (orange arrow) become more pronounced compared with the hydrodynamical modes (purple arrows) predicted by Eq.~\ref{Sfrequency}. 
}
\label{StoBtransition}
\end{figure}

We apply a weak quench which changes the interaction strength from $g_0$ to $g_1$ where $g_1\equiv g_0g_x$. Conventionally, we examine the low-lying excitations by looking at the oscillatory behavior of particle density $\rho({\bf{r}},t)=|\psi({\bf{r}},t)|^2$ \cite{PhysRevLett.77.2360}. Due to cylindrical symmetry, the particle density only depends on $r$, and we can replace $\rho(\mathbf{r},t)$ by  $\rho(r,t)$. We take the Fast Fourier Transform (FFT) of the density from the time domain to frequency domain, denoted as $\rho(r,\omega_n)$. At short length scales where $r$ is small, we find that, compared with the excitations predicted by conformal symmetry (i.e., $\omega_n = 2n\omega_0$), additional excitations emerge that signal a breakdown of scale invariance. 
Intriguingly, these additional modes can be accurately described by Eq.~\ref{Sfrequency}, derived from a perturbative analysis within the hydrodynamic theory of a 2D BEC~\cite{RevModPhys.81.647, PhysRevLett.77.2360, LuoChen2013}: 
\begin{equation}\label{Sfrequency}
\omega_n = \omega_0 \left( 2n^2 + 2nl+ 2n +l\right)^{1/2}.
\end{equation}
In particular, for the lowest four modes (which have zero angular momentum $l$), our numerical results for $\omega_n/\omega_0$ (1.98, 3.49, 4.94, 6.38) show good agreement with the theoretical predictions (2.00, 3.46, 4.90, 6.32) from Eq.~\ref{Sfrequency}, as detailed in Fig.~\ref{Stringari_mode}. 
When the observational length scale $r$ is large, or if the quench becomes stronger
(at higher $g_x$), we find more pronounced even-integer modes (orange arrows), as indicated in Fig.~\ref{StoBtransition}.
In this case, a contact potential approach is a good approximation as the finite cut-off size of the system is negligible, so that the conformal symmetry is effectively restored.

\section{Dispersion relation of high-lying excitations}\label{QSection}

When studying the high-energy excitations, it is difficult to address their frequencies in real space as they 
may coexist with low-lying modes while having a much weaker strength. Momentum space variables are more
appropriate as we observe that their frequency scales with momentum. This is in contrast to the low-lying excitations, which only depend on the trap frequency $\omega_0$. This reflects the sound-mode nature of the high-lying excitations characterized by the Bogoliubov theory. These modes with higher momenta are restricted to short distances in real space, and thus are less affected by the trap effects. 
They are more readily characterized through the wavefunction perturbation $\delta\psi$ around the ground state $\psi_0$ at $g_0$, where $\psi_0$ is real and takes the Thomas-Fermi form \cite{PhysRevLett.77.2360}. The time evolution of $\delta\psi$ is governed by the linearized GP equation in Eq.~\ref{deltapsipde}:
\begin{equation}\label{deltapsipde}
\begin{aligned}
&i\hbar\frac{\partial\delta\psi(\mathbf{r},t)}{\partial t}
=-\frac{\hbar^{2}\nabla^{2}}{2m}\delta\psi(\mathbf{r},t)
+ g_0\psi_0^2
\left[\delta\psi(\mathbf{r},t)+\delta\psi^{*}(\mathbf{r},t)\right] \\
&=-\frac{\hbar^{2}\nabla^{2}}{2m}\delta\psi(\mathbf{r},t)
+\left(\mu-\frac{1}{2}m\omega_0^{2}r^{2}\right)
\left[\delta\psi(\mathbf{r},t)+\delta\psi^{*}(\mathbf{r},t)\right].
\end{aligned}
\end{equation}

\begin{figure}[h]
\vspace{0pt}
\centering
\includegraphics[width=1\linewidth]{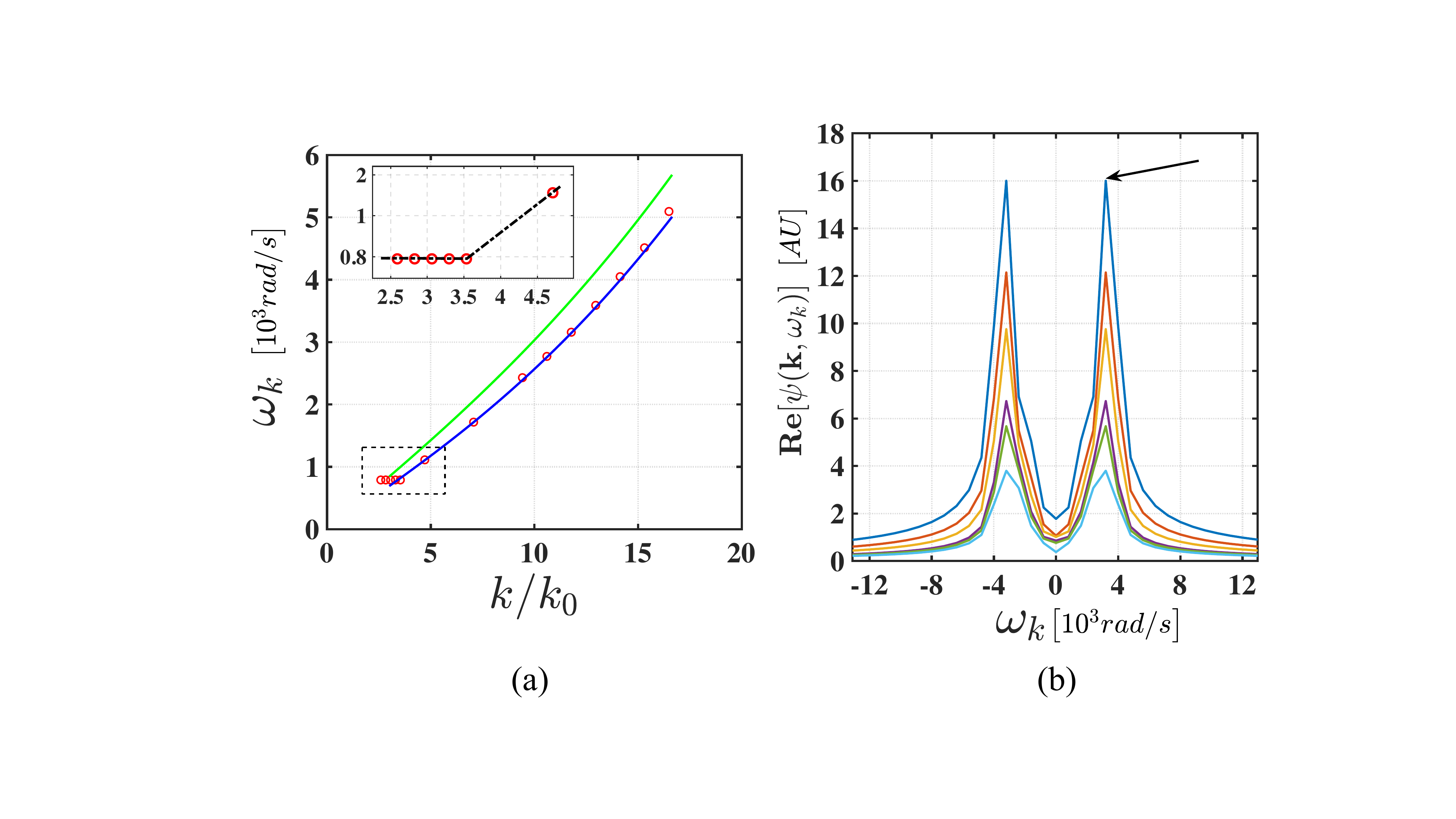}
\caption{
\textbf{(a)} At a fixed trapping frequency $\omega_0=60~\mathrm{rad/s}$, 
the predictions from Bogoliubov theory with $\mu_{\text{eff}}$ 
(blue line) agree with the numerical data (red dots) within a reasonable error from $0.41\%$ to $2.95\%$.
By contrast, the standard Bogoliubov theory (green line), which assumes a homogeneous density distribution, fails to reproduce these oscillation frequencies.
For small wavenumbers ($k<4k_0$), 
the excitation frequencies saturate to a constant value independent of $k$, indicating a crossover to low-lying collective modes.\\
\textbf{(b)} 
When excitations with a given $\mathbf{k} = (0, 2.6)\mu\mathrm{m}^{-1}$ are measured under different trapping frequencies $\omega_0$ (ranging from $60~\mathrm{rad/s}$ to $125~\mathrm{rad/s}$), their excitation frequencies remain unchanged once the modes enter the large-$k$ regime. Meanwhile, the peak amplitudes $\mathrm{Re}\left[\psi(\mathbf{k},\omega_k)\right]$ decrease as the trapping frequency increases.}
\label{swfrequencies}
\end{figure}

Numerically, this wavefunction perturbation $\delta\psi$ is accessible. We perform a FFT on  $\delta\psi(\mathbf{r},t)$ to obtain $\delta\psi(\mathbf{k},t)$ and then investigate its oscillatory dependence over time. 
Figure~\ref{swfrequencies} shows that, importantly, these high-$k$ sound modes in a trapped
configuration can be described by the usual Bogoliubov theory with a modified chemical potential $\mu_{\text{eff}}$ given as
\begin{equation}\label{mueff}
\mu_{\text{eff}}\equiv
\frac{
\int_0^{r_0} |\psi_0(r)|^2
\left[
\mu-\frac{1}{2}m\omega_0^{2}r^{2}\right]
rdr
}{\int_0^{r_0} |\psi_0(r)|^2rdr}
\approx\frac{2\mu}{3},
\end{equation}
where the Thomas-Fermi radius $r_0$ describes the boundary of the condensate and a characteristic wavevector can be defined according to this length scale as $k_0 = 2\pi/r_0$. Fig.~\ref{swfAgainstw} indicates this numerical procedure. The analytical derivations are shown in Appendix.~\ref{sec:dispersion}.

\begin{figure}[h]
\vspace{0pt}
\centering
\includegraphics[width=\linewidth]{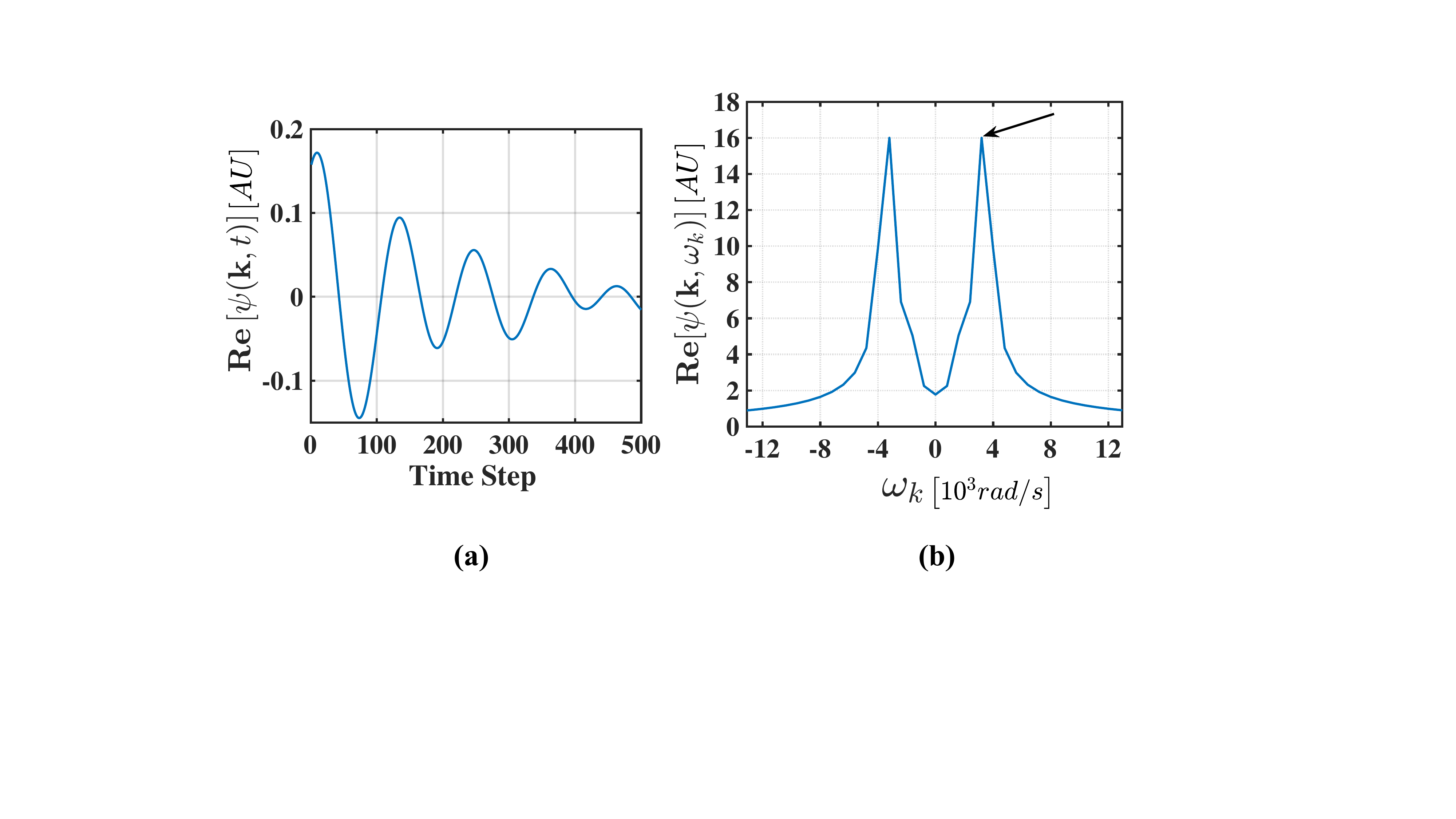}
\caption{
\textbf{(a)} An example of the oscillation of $\operatorname{Re}[\psi(\mathbf{k},t)]$ over time at $(k_x,k_y) = (0,\,2.6)\,\mu\mathrm{m}^{-1}$, from which the excitation frequency is extracted. The unit for each computational time step is $16~\mu\mathrm{s}$.\\
\textbf{(b)} The corresponding Fourier transform of the signal shown in (a).}
\label{swfAgainstw}
\end{figure}

Experimentally, the high-lying excitations in trapped Bose Einstein condensates (BECs) are
probed via the dynamic structure factor under a quench protocol:
\begin{equation*} 
S_q(\mathbf k,t)=\frac{\langle|\Delta\rho(\mathbf k,t)|^{2}\rangle}{N}, 
\end{equation*} 
where $\Delta\rho(\mathbf k,t)$ is the Fourier component of the density deviation and $N$ is the total particle number.  

\begin{figure}[h]
\vspace{0pt}
\centering
\includegraphics[width=0.65\linewidth]{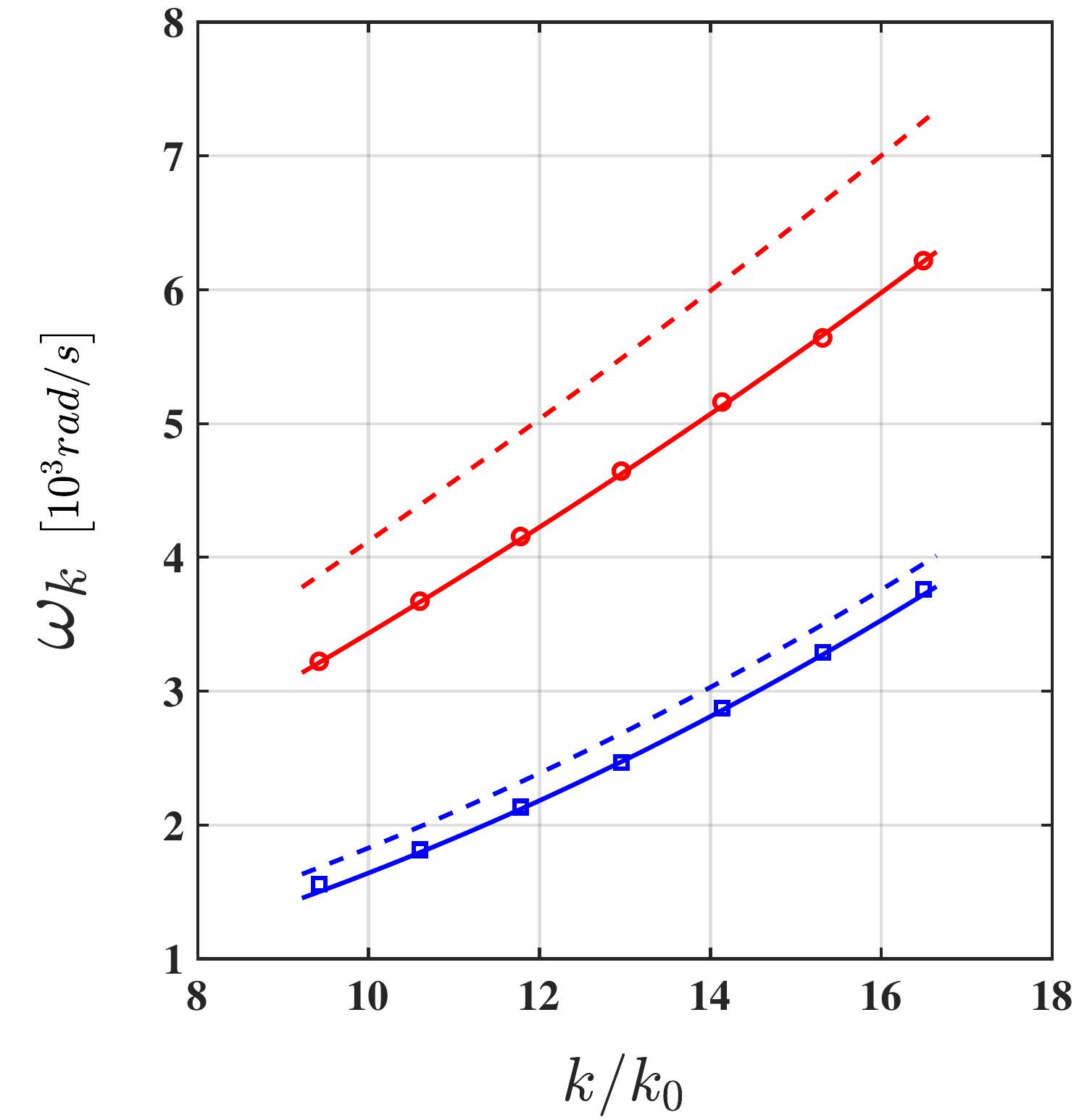}
\caption{
This figure shows good agreement between the modified Bogoliubov theory given in Eq.~\ref{dispersionquench} (solid lines) and the  numerical data (hollow symbols) of the excitation frequencies extracted from $S_q(\mathbf{k},t)$ for various $k$ (from $9\,k_0$ to $17\,k_0$). We show both quench up (red) at $g_0 = 3$ with $g_x = 2$ and quench down (blue) at $g_0 = 13$ and $g_x = 0.25$ cases. The Bogoliubov theory without trap effects (dashed lines) fails to capture these oscillation frequencies as expected. We extracted the corresponding frequencies from the first five oscillations of structure factors. }
\label{quenchedcases}
\end{figure}

The structure factor is predicted to oscillate at twice the frequency of the sound mode at $\bf{k}$ (analytic details are presented in Appendix~\ref{sec:Sq}). So we extract the excitation frequencies from $S_q(\mathbf{k},t)$ and  present our numerical results under both quench-up and quench-down conditions in Figure~\ref{quenchedcases}. The agreement between the predictions of our theoretical model (solid lines) given by Eq.~\ref{dispersionquench} and our numerical data (hollow symbols) across both scenarios is quite reasonable. It is particularly noteworthy that for the quench-up case, standard Bogoliubov theory (dashed lines), assuming a homogeneous density, yields frequencies that significantly deviate from our numerical observation, a discrepancy that is more pronounced than in the quench-down case. 

This discrepancy~\cite{ChengChin}, can now
be removed by switching to the effective ``chemical potential" $\tilde\mu_{\text{eff}}=g_x\mu_{\text{eff}}$ corresponding to the interaction strength after quench. 
In our study, in contrast to the local
density approximation where the chemical potential is position dependent,
we find that it is appropriate to define a renormalized global chemical potential, $\mu_{\mathrm{eff}}$ which inherently incorporates contributions from the external trapping potential.
Our numerics demonstrates that this provides a good approximation in the high-$k$ regime. 

As follows from Eq.~\ref{dispersionquench}, using the standard Bogoliubov theory is more problematic for the quench-up process as it probes
the regime where the influence of the linear term in the dispersion relation (which depends on $\tilde\mu_{\text{eff}}$) is substantial. In contrast, the quench-down case primarily focuses on the quadratic regime, where this linear 
term is less consequential. These findings highlight the critical importance of incorporating trap effects, as we do 
here, for a reliable description of collective excitations in realistic trapped BECs.

Our numerical results also reveal a noticeable decay of these high-$k$ oscillations over time. This decay arises because these excitations are only approximately eigenmodes of momentum in presence of the translation-invariance breaking trap potential. 
Consequently, an initially seeded excitation at momentum $\mathbf{k}$ disperses into various eigenmodes.
Physically, the seeded excitation gradually propagates out of the condensate region; once it exits, the oscillation amplitude diminishes and eventually vanishes.

\begin{figure}[H]
\vspace{0pt}
\centering
\includegraphics[width=\linewidth]{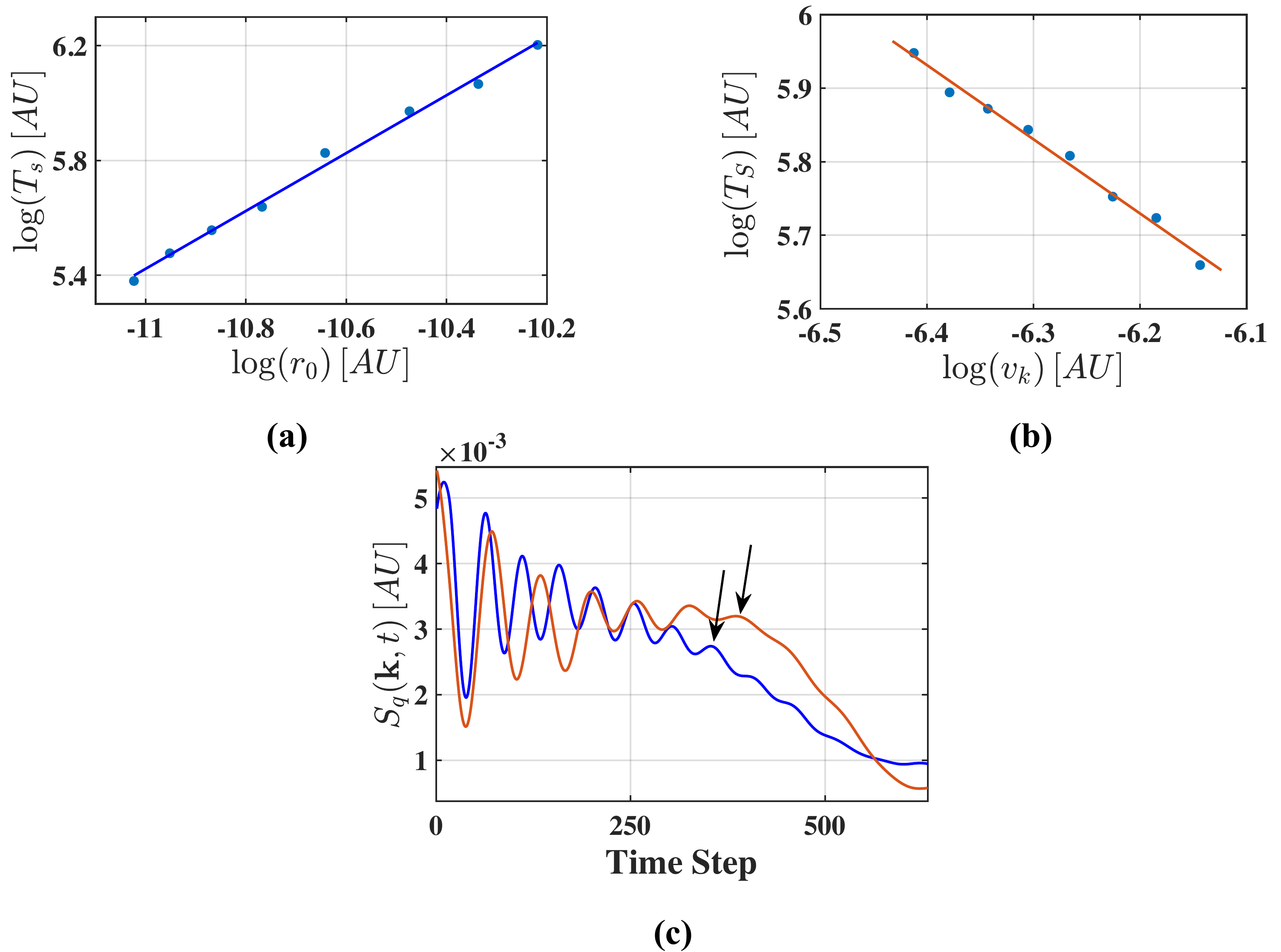}
\caption{
Both (a) and (b) shows the scaling behaviors of stabilization time illustrating the validity of Eq.~\ref{Ts}. 
\textbf{(a)}
Varying the trap frequency $\omega_0$ from $49~\mathrm{rad/s}$ to $109~\mathrm{rad/s}$ to effectively change the condensate size $r_0$, while examining at $k = 2.6\,\mu\mathrm{m}^{-1}$ in Fourier space, yields the numerical results shown as blue dots. A linear regression (blue solid line) gives $T_S \propto r_0^{1.0}$.
\textbf{(b)} Varying $k$ from $8\,k_0$ to $17\,k_0$ to vary the group velocity $v_k$ while keeping $\omega_0$ fixed to be $60$ ad/s, which shows that a linear regression (red solid line) of the numerical results (blue dots) yields
$T_S \propto v_k^{-1.0}$. 
\textbf{(c)} Illustration of taking $T_s$ (denoted by black arrows) for $(k,\omega_0) = (11k_0,60~\mathrm{rad/s})$ (blue line) and $(8.3k_0,63~\mathrm{rad/s})$ (red line).
}
\label{Tsvsk}
\end{figure}

Based on these considerations, this provides an estimate of the lifetime $T_s$:
\begin{equation}\label{Ts}
T_{s}= \frac{r_0}{v_k},
\end{equation}
where $v_k = d\omega_k/dk$ is the group velocity derived from the oscillation frequency $\omega_k$ as given by Eq.~\ref{dispersionquench}. In Figure~\ref{Tsvsk}, we show that our numerical results for this lifetime generally align with this theoretical prediction for a range of different parameters.

\newpage

\section{Conclusion}

This work presented a comprehensive analysis of the collective modes excited by interaction quenches in a harmonically trapped two-dimensional (2D) Bose-Einstein condensate (BEC), focusing on both the low and high-lying excitation regimes in a more realistic and non-idealized situation.

Crucially, our study unveiled a complex collection of oscillatory modes in the low-momentum regime which
arise in the presence of a quench. This corresponds to a hybrid collective mode structure, composed of excitations characteristic of hydrodynamic modes \cite{PhysRevLett.77.2360} as well as even-integer modes predicted in
Ref.~\cite{PhysRevA.55.R853}. This complexity reflects the breakdown of the underlying SO(2,1) conformal symmetry at short observational distances. In this way one finds a form of hybrid collective mode spectrum at low energies.
We argue it should be
accessible and, thus, testable in future experiments.

When probing higher momentum, the excitations become momentum dependent and follow a modified Bogoliubov dispersion relation which now contains a renormalized chemical potential $\mu_{\text{eff}}$. 
Also relevant is the behavior of the finite lifetime of these high-k modes which we demonstrate can be
understood.
This modification of Bogoliubov theory derives from the effects of the harmonic trap. We suggest that it may offer
insights into theoretical discrepancies~\cite{ChengChin} observed in quench-up scenarios when compared with homogeneous theory.

\section*{Acknowledgments}
A.G. was supported by the U.S. Department of Energy, Office of Science, Basic Energy Sciences, Materials Sciences and Engineering Division. We also acknowledge the University of Chicago's Research Computing Center for their support of this work.

\bibliography{reference}

\newpage
\appendix

\section{Dispersion Relations}\label{sec:dispersion}


To simplify Eq.\ref{deltapsipde} for high-lying excitations with big enough $\vec{k}$ comparing with $|\vec{k}_0| = 2\pi/r_0$, we exploit a separation of length scales. This allows us to approximate the spatially varying trapping potential by its average effect over the ground state. We average the potential term over the ground-state density, yielding an effective constant $\mu_{\text{eff}}$ given in Eq.~\ref{mueff}.
This approximation simplifies Eq.~\ref{deltapsipde} to an effective homogeneous Bogoliubov equation for $\delta\psi(\mathbf{r},t)$:
\begin{equation} \label{perturbednGP}
i\hbar\frac{\partial\delta\psi(\mathbf{r},t)}{\partial t}
=-\frac{\hbar^{2}\nabla^{2}}{2m}\delta\psi(\mathbf{r},t)
+\mu_{\text{eff}}
\left(\delta\psi(\mathbf{r},t)+\delta\psi^{*}(\mathbf{r},t)\right),
\end{equation}
where $\mu_{\text{eff}} = 2\mu/3$ acts as an renormalized chemical potential that incorporates the average influence of the trap.

Assuming plane-wave solutions of the form $\delta\psi(\mathbf{r},t)=\frac{1}{\sqrt{V}}\sum_{\mathbf{k}}(u_{\mathbf{k}}e^{-i(\mathbf{k}.\mathbf{r}-\omega_{k}t)}+v_{\mathbf{k}}e^{i(\mathbf{k}.\mathbf{r}-\omega_{k}t)})$, where $V$ is the system volume, Eq.~\ref{perturbednGP} leads directly to the Bogoliubov dispersion relation, but with the renormalized chemical potential $\mu_{\text{eff}}$:
\begin{equation}\label{dispersion}
\omega_{k}
=\sqrt{\frac{\mu_{\text{eff}}}{m}k^{2}+\frac{\hbar^{2}}{4m^{2}}k^{4}}
=vk\sqrt{1+\left(\frac{\hbar k }{2 m v}\right)^2},
\end{equation}
where $v^2 = \mu_{\text{eff}}/m$ represents the renormalized speed of sound.

To further analyze the form of the perturbation $\delta\psi(\mathbf r,t)$, the coefficients $u_{\mathbf{k}}$ and $v_{\mathbf{k}}$ can be related through the Bogoliubov transformation parameters $\cosh(\alpha_{k})$ and $\sinh(\alpha_{k})$:
\begin{equation}
\begin{aligned}
\left(\begin{array}{c}
u_{\mathbf k}\\
v_{\mathbf k}^{*}
\end{array}\right)
=
\left(\begin{array}{cc}
-\cosh(\alpha_{k}) & -\sinh(\alpha_{k})\\
\sinh(\alpha_{k}) & \cosh(\alpha_{k})
\end{array}
\right)
\left(\begin{array}{c}
u_{\mathbf k}\\
v_{\mathbf k}^{*}
\end{array}
\right),
\end{aligned}
\end{equation}
where $\cosh(\alpha_{k})=\frac{\hbar^{2}k^{2}}{2m\varepsilon_{k}}+\frac{\mu_{\text{eff}}}{\varepsilon_{k}}$ and $\sinh(\alpha_{k})=\frac{\mu_{\text{eff}}}{\varepsilon_{k}}$, with $\varepsilon_{k}=\hbar\omega_{k}$. This allows rewriting $\delta\psi(\mathbf r,t)$ in a compact form:
\begin{equation}\label{deltapsiAQ}
\begin{aligned}
\frac{1}{\sqrt{V}}
&\sum_{\mathbf{k}}
u_{\mathbf{k}}e^{-i\left(\mathbf{k}.\mathbf{r}-\omega_{k}t\right)}
-\coth\left(\frac{\alpha_{k}}{2}\right)u_{\mathbf{k}}^{*}
e^{i(\mathbf{k}.\mathbf{r}-\omega_kt)}.
\end{aligned}
\end{equation}

Next, we consider the system immediately after a quench, where the interaction strength changes from $g_0$ to $g_1$. A similar reduced GP equation holds for the new weak perturbation $\Delta\psi$ around the post-quench ground state $\psi_1$. We express $\Delta\psi$ in an analogous form:
\begin{equation}\label{DeltapsiAQ}
\begin{aligned}
&\frac{1}{\sqrt{V}}\sum_{\mathbf{k}}
U_{\mathbf{k}}e^{-i\left(\mathbf{k}.\mathbf{r}-\Omega_{k}t\right)}
-\coth\left(\frac{\beta_{k}}{2}\right)U_{\mathbf{k}}^{*}
e^{i(\mathbf{k}.\mathbf{r}-\Omega_kt)},
\end{aligned}
\end{equation}
where $\cosh(\beta_{k})\equiv\frac{\hbar^{2}k^{2}}{2mE_{k}}+\frac{\tilde\mu_{\text{eff}}}{E_{k}}$, $\sinh(\beta_{k})\equiv\frac{\tilde\mu_{\text{eff}}}{E_{k}}$ and $E_k = \hbar\Omega_k$. This implies a new dispersion relation for $\Delta\psi(\mathbf r,t)$ with an renormalized chemical potential $\tilde{\mu}_{\text{eff}}\equiv g_x\mu_{\text{eff}} $:
\begin{equation}\label{dispersionquench}
\Omega_{k} = \sqrt{\frac{\tilde{\mu}_{\text{eff}}}{m}k^{2}+\frac{\hbar^{2}}{4m^{2}}k^{4}}.
\end{equation}

\section{Structure factor}\label{sec:Sq}

Based on the definition of the density deviation $\delta\rho(\mathbf  r,t)\equiv\rho( \mathbf  r,t)-\rho_{0}(r)$, where $\rho_{0}(r)$ is the time average of $\rho(\mathbf r,t)$, one can work out $\delta\rho(\mathbf  r,t)$ in real space as $\psi_{0}(r)\bigl[\delta\psi(\mathbf r,t) +\delta\psi^{*}(\mathbf  r,t)\bigr]$. Since we consider $\delta\psi(\mathbf{r},t)$ in a separate length scale much smaller than $r_0$, the Fourier transform of $\delta\rho(\mathbf  r,t)$ can be approximated as
\begin{equation}
    \begin{aligned}
        \delta\rho(\mathbf{k},t)\approx& \sqrt{n_0}
     \bigl[\delta\psi(\mathbf{k},t)+\delta\psi^{*}(\mathbf{k},t)\bigr],
     \text{ where } n_0 =\frac{N}{V}
   \\
    \delta\psi(\mathbf{k},t) =& u_{-\mathbf{k}}e^{i\omega_k t}-u_{\mathbf{k}}^* \coth\left(\frac{\alpha_{k}}{2}\right) e^{-i\omega_k t}
    \end{aligned}
\end{equation}
Therefore, for a given $\mathbf k$, 
\begin{equation}
    \begin{aligned}
        \delta\rho(\mathbf k,t)=
        & \sqrt{n_0}\,[\,\tilde u_{-\mathbf k}(t)+ \tilde u_{\mathbf k}^{*}(t)\,],
    \end{aligned}
\end{equation}
where $ \tilde u_{-\mathbf k}(t)\equiv  u_{-\mathbf k}e^{i\omega_{k}t}\left(1-\coth\left(\alpha_{k}/2\right)\right)$. Thus, one can further compute $\left|\delta\rho(\mathbf k,t)\right|^2$ and take the statistical average to obtain $S\left(\mathbf k\right) $. The Bose-Einstein statistics gives $\left\langle |u_{\mathbf k}|^{2}\right\rangle/V = \left\langle |u_{-\mathbf k}|^{2}\right\rangle /V\equiv 1/(e^{\beta\varepsilon_{k}}-1)$ for a given temperature
, and $\left\langle u_{\mathbf k}^{2}\right\rangle =\left\langle u_{\mathbf k}u_{-\mathbf{k}}\right\rangle= \left\langle u_{-\mathbf k}^{*2}\right\rangle = 0$ for a given $\mathbf{k}$ at finite temperature \cite{bose1924planck,einstein1924quantentheorie}.

The structure factor without the quench for the sound modes with large $k$ can be derived as the following 
\begin{equation}\label{sfbeforequench}
\begin{aligned}
    S\left(\mathbf k\right) =&\frac{\langle|\delta\rho(\mathbf k,t)|^{2}\rangle}{N}
= 
\frac{\langle|u_{\kappa}|^{2}\rangle+\langle|u_{-\kappa}|^{2}\rangle}{V}
\left[1-\coth\left(\frac{\alpha_{\kappa}}{2}\right)\right]^{2}\\
=&\bigg[
\frac{2}{e^{\beta\varepsilon_{k}}-1}
\bigg]\frac{e^{-\alpha_{\kappa}}}{\sinh^{2}\left(\alpha_{\kappa}/2\right)}\\
=&
\frac{2}{\sinh^{2}\left(\alpha_{\kappa}/2\right)}\frac{\hbar^{2}k^{2}}{2m\varepsilon_{k}}\left[\frac{1}{e^{\beta\varepsilon_{k}}-1}\right]. 
\end{aligned}
\end{equation}

To connect the pre- and post-quench dynamics and derive the post-quench structure factor $S_q(k,t)$, we invoke the continuity of the wavefunction at $t=0$:
\begin{equation}
\label{continuity}
\begin{aligned}
&\psi_0(r)+\delta\psi(\mathbf r,t=0)=\psi_1(r)+\Delta\psi(\mathbf r,t=0)\\
\Rightarrow
&\delta\psi(\mathbf k,t=0)=\Delta\psi(\mathbf k,t=0), \forall k\gg k_0.
\end{aligned}
\end{equation}
This continuity condition 
gives 
\begin{equation}
    \begin{aligned}
        u_{-\mathbf{k}}-\coth\left(\frac{\alpha_{{k}}}{2}\right)u_{\mathbf{k}}^{*}
        =&U_{-\mathbf{k}}-\coth\left(\frac{\beta_{k}}{2}\right)U_{\mathbf{k}}^{*}\\
      u_{\mathbf{k}}^*-\coth\left(\frac{\alpha_{{k}}}{2}\right)u_{-\mathbf{k}}        =&U_{\mathbf{k}}^*-\coth\left(\frac{\beta_{k}}{2}\right)U_{-\mathbf{k}},\\
\end{aligned}
\end{equation}
which allows us to express $U_{-\mathbf{k}}$ in terms of $u_{-\mathbf{k}}$ and $u_\mathbf{k}^*$ as 
\begin{equation}
    \begin{aligned}
        \frac{\sinh\left(\beta_{k}/2\right)}{\sinh\left(\alpha_{k}/{2}\right)}
        \Bigg[u_{-\mathbf{k}}\cosh\left(\frac{\beta_{k}-\alpha_{k}}{2}\right)+
        u_{\mathbf{k}}^{*}\sinh\left(\frac{\beta_{k}-\alpha_{k}}{2}\right)\Bigg].
    \end{aligned}
\end{equation}
Therefore, one can easily work out the following expressions,
\begin{equation}
    \begin{aligned}
        \left\langle |U_{\mathbf{k}}|^{2}\right\rangle 
        =&\left[\frac{\sinh\left(\beta_{k}/2\right)}{\sinh\left(\alpha_{k}/2\right)}\right]^{2}
        \cosh\left(\beta_{k}-\alpha_{k}\right)\langle |u_\mathbf{k}|^{2}\rangle \\
        \langle U_{-\mathbf{k}}U_\mathbf{k}\rangle 
        =&\left[\frac{\sinh\left(\beta_{k}/2\right)}{\sinh\left(\alpha_{k}/2\right)}\right]^{2}
        \sinh\left(\beta_{k}-\alpha_{k}\right)\langle \left|u_\mathbf{k}\right|^{2}\rangle. 
    \end{aligned}
\end{equation}
Notably, $\langle U_{-\mathbf{k}}U_\mathbf{k}\rangle $ and $\langle U^*_{-\mathbf{k}}U^*_\mathbf{k}\rangle $ no longer vanish.
Consequently, the structural factor after the quench can be derived as
\begin{equation}
    \begin{aligned}
        &S_q(\mathbf{k},t)=\frac{\langle \left|\Delta\rho(\mathbf{k},t)\right|^{2}\rangle}{N}=\left[1-\coth\left(\frac{\beta_{\kappa}}{2}\right)\right]^{2} \times\\
        &\frac{\langle|U_\mathbf{k}|\rangle^{2}+\langle|U_{-\mathbf{k}}|\rangle^{2}
        +\langle U_\mathbf{k}U_{-\mathbf{k}}\rangle e^{2i\Omega_{\kappa}t}
        +\langle U_{\mathbf{k}}^{*}U_{-\mathbf{k}}^{*}\rangle e^{-2i\Omega_{\kappa}t}}{V}
        \\
&=S(\mathbf{k})\left[\frac{E_{k}^{2}+\varepsilon_{k}^{2}}{2E_{k}^{2}}+\frac{E_{k}^{2}-\varepsilon_{k}^{2}}{2E_{k}^{2}}\cos\left(2\Omega_{k}t\right)\right] \\
        &=S(\mathbf{k})\left[1-\frac{E_{k}^{2}-\varepsilon_{k}^{2}}{E_{k}^{2}}\sin^{2}\left(\Omega_{k}t\right)\right],
    \end{aligned}
\end{equation}
where $S(\mathbf{k})$ is the initial structure factor given by Eq.~\ref{sfbeforequench}, $\Delta\rho(\mathbf{k},t)$ is the density deviation after the quench, and $\Omega_k$ is defined in Eq.~\ref{dispersionquench}. This expression for $S_{q}(\mathbf{k},t)$ is consistent with derivations from density-density correlation functions by Hung \textit{et al.}~\cite{ChengChin}.

\section{Numerical seeding}\label{SeedingQ}

Due to the fact that quantum fluctuations are not naturally encoded in the GP equation, we simulate the fluctuations at certain $\bf{k}$ by adding seeds.

To study the low-lying excitations, we introduced a seed that rigorously preserves cylindrical symmetry (carrying zero angular momentum), $\psi(r,t=0)\;\rightarrow\;\psi_0(r)\bigl[1+A\,e^{i k r}\bigr]$. The seed is implemented at $k= 0.26~\mu m^{-1}$ and the amplitude A is taken to be 0.01 in our simulations.

For the high-lying excitations, when studying them by directly looking into the oscillation of Re$\left[\psi( \mathbf{k}, t )\right]$ without a quench, we implemented a weak single-frequency wave packet as a noise around the ground state such that $\psi(x,y) = \psi_0(x,y)(1+A e^{ik_x x +ik_y y})$, at $\mu/\hbar\omega_0=52.51$. We fix $k_x=0$ and vary $k_y$ between $2.59k_0$ and $16.49k_0$ and look into the spectrum of Re$\left[\psi( \mathbf{k}, \omega_k )\right]$ in frequency space. 
A is also taken to be 0.01 here. 

When studying the high-lying excitations with the quench, we intend to simulate the structure factor as physical as possible. There are several ingredients to consider. 
First of all, instead of a single seed, we apply seeds to all momenta at the same $|k|$ to simulate Eq.~\ref{deltapsiAQ}. In order to recover Bose-Einstein statistics, for each pair of $\bf{k}$ and $-{\bf{k}}$, we explicitly introduce two randomly generated phase factor $\varphi_1$ and $\varphi_2$, such that $u_{\mathbf k} \equiv |u_{\mathbf k}| e^{i\varphi_1}$ and $u_{-\mathbf k}\equiv |u_{\mathbf k}| e^{i\varphi_2}$. The Bose-Einstein statistics can be ensured because $|u_{\mathbf k}|^2$ corresponds to $1/(e^{\beta\varepsilon_{k}}-1)$ for certain $\beta$  and  $\left\langle u_{k}^{2}\right\rangle = |u_{\mathbf k}|^{2}\left\langle  e^{-2i\varphi_1}\right\rangle = 0$. 
Since these seeds are simulating quantum fluctuations prior to quench, coefficients $\cosh(\alpha_k)$ and $\sinh(\alpha_k)$ are calculated using the ground state parameters (with coupling constant equal to $g_0$). 
Secondly, at $t=0$, the total perturbation $\delta\psi({\bf{k}})$ at ${\bf{k}}$ is $u_{\mathbf k}-u^*_{-\mathbf k}\coth(\alpha_k/2)$ and for $-{\bf{k}}$, it is $u_{-\mathbf k}-u^*_{\mathbf k}\coth(\alpha_k/2)$. To be compatible with the ground state $\psi_0$ which has finite size in real space, we add back these seeds to the ground state as 
$\psi({\bf{r}})=\psi_0({\bf{r}})\left(1+\sum_{{\bf{k}}}\delta\psi({\bf{k}})e^{i{\bf{k}}\cdot{\bf{r}}}\right)$. Here, 
$|u_{\mathbf{k}}|$ is fixed to be $0.8\times10^{-3}$ throughout all directions, and $\varphi_1$ and $\varphi_2$ are repeatedly and randomly generated for each pair of seeds implemented in the opposite direction. We add 10 evenly spaced plane-wave seeds with non-zero $k_x$ and $k_y$ in $k$-space, with a fixed $k=\sqrt{k_x^2+k_y^2}.$

\end{document}